\documentclass[letterpaper,english,american,twocolumn,aps,prb,showpacs,superscriptaddress]{revtex4-1}
\usepackage[T1]{fontenc}
\usepackage[latin9]{inputenc}
\usepackage{color}
\usepackage{textcomp}
\usepackage{amsmath}
\usepackage{amssymb}
\usepackage{graphicx}

\makeatletter

\usepackage{ulem}
\usepackage{float}
\usepackage{lipsum}
\usepackage{babel}

\makeatother

\usepackage{babel}
\begin{document}
\title{Dynamical singularity of the rate function for quench dynamics in
finite-size quantum systems}
\author{Yumeng Zeng}
\affiliation{Beijing National Laboratory for Condensed Matter Physics, Institute
of Physics, Chinese Academy of Sciences, Beijing 100190, China}
\affiliation{School of Physical Sciences, University of Chinese Academy of Sciences,
Beijing 100049, China }
\author{Bozhen Zhou}
\affiliation{Beijing National Laboratory for Condensed Matter Physics, Institute
of Physics, Chinese Academy of Sciences, Beijing 100190, China}
\author{Shu Chen}
\email{Corresponding author: schen@iphy.ac.cn }

\affiliation{Beijing National Laboratory for Condensed Matter Physics, Institute
of Physics, Chinese Academy of Sciences, Beijing 100190, China}
\affiliation{School of Physical Sciences, University of Chinese Academy of Sciences,
Beijing 100049, China }
\affiliation{Yangtze River Delta Physics Research Center, Liyang, Jiangsu 213300,
China }
\date{\today}
\begin{abstract}
The dynamical quantum phase transition is characterized by the emergence
of nonanalytic behaviors in the rate function, corresponding to the
occurrence of exact zero points of the Loschmidt echo in the thermodynamical
limit. In general, exact zeros of the Loschmidt echo are not accessible
in a finite-size quantum system except for some fine-tuned quench
parameters. In this work, we study the realization of the dynamical
singularity of the rate function for finite-size systems under the
twist boundary condition, which can be introduced by applying a magnetic
flux. By tuning the magnetic flux, we illustrate that exact zeros
of the Loschmidt echo can be always achieved when the postquench parameter
is across the underlying equilibrium phase transition point, and thus
the rate function of a finite-size system is divergent at a series
of critical times. We demonstrate our theoretical scheme by calculating
the Su-Schrieffer-Heeger model and the Creutz model in detail and
exhibit its applicability to more general cases. Our result unveils
that the emergence of dynamical singularity in the rate function can
be viewed as a signature for detecting dynamical quantum phase transition
in finite-size systems. We also unveil that the critical times in
our theoretical scheme are independent on the systems size, and thus
it provides a convenient way to determine the critical times by tuning
the magnetic flux to achieve the dynamical singularity of the rate
function. 
\end{abstract}
\maketitle

\section{Introduction}

Since the dynamical quantum phase transition (DQPT) was proposed \citep{Heyl2013PRL},
it has become an important concept in describing a class of nonequilibrium
critical phenomena associated with singular behavior in the real-time
evolution of the Loschmidt echo (LE) \citep{Heyl2013PRL,Karrasch2013PRB,Hickey2014PRB,Andraschko2014PRB,Schmitt2015PRB,Vajna2014PRB,Sharma2015PRB,Heyl2018RPP,Canovi2014PRL,
Mera2018PRB,Zvyagin2016LTP,Heyl2018RPP,YangC,Budich2016PRB,Heyl2015PRL,Heyl2014PRL,Zhou2019PRB,BoBo2020PRB,Halimeh,Dora,Halimeh2,Jafari2}. Given $\langle\psi_{i}|\psi(t)\rangle$ denotes the overlap of an
initial ground state $|\psi_{i}\rangle$ and its time evolution state
$|\psi(t)\rangle=e^{-iH_{f}t}|\psi_{i}\rangle$ governed by a postquench
Hamiltonian $H_{f}$, the LE is defined as 
\begin{equation}
\mathcal{L}(t)=|\langle\psi_{i}|\psi(t)\rangle|^{2},
\end{equation}
which represents the return probability of the time evolution state
to the initial state \citep{Gorin2006PR}. The LE plays a particularly
important role in the characterization of the DQPT \citep{Zvyagin2016LTP,Heyl2018RPP}.
When the phase-driving parameter is quenched across an underlying
equilibrium phase transition point, a series of zero points of LE
emerge at some critical times. In general, exact zeros of LE only
occur when the system size tends to infinity \citep{Heyl2013PRL,Liska,ZhouBZ2021}.
Meanwhile, LE always approaches zero in the thermodynamical limit,
even when the quench parameter does not cross the transition point.
This can be attributed to the Anderson orthogonality catastrophe \cite{OC}
for the reason that the multiplication of an infinite number of numbers
with magnitude less than 1 equals 0. To eliminate the effect of system
size properly, it is convenient to introduce the rate function of
LE given by 
\begin{equation}
\lambda(t)=-\frac{1}{L}\ln\mathcal{L}(t).\label{rf}
\end{equation}
As the LE is analogous to a dynamical boundary partition function,
the rate function $\lambda(t)$ can be viewed as a dynamical free
energy.
Thus the DQPT is characterized by nonanalytic behaviors in the rate
function of LE in the thermodynamical limit. 

According to the theory of DQPT, the nonanalyticity of rate function
occurs at the critical times $t_{n}^{*}$ when the quench parameter
is across the equilibrium phase transition point, corresponding to
the emergence of exact zeros of LE in the thermodynamical limit. For
finite size systems, LE usually has no exact zeros, except for fine-tuned
post-quench parameters which fulfill specific constraint conditions
\cite{ZhouBZ2021}. Therefore, to study the DQPT and extract the critical
times in finite-size quantum systems, one needs to resort to finite-size
analysis to extract the non-analytical properties and critical times
in the limit of $L\rightarrow\infty$. With the increase of $L$,
$\mathcal{L}(t)$ approaches zero at critical times $t_{n}^{*}$,
and thus $\ln\mathcal{L}(t_{n}^{*})\rightarrow\infty$ when $L\rightarrow\infty$.
However, $\lambda(t_{n}^{*})$ is not divergent and only displays
a cusp due to the fact that the divergence is offset by the $L$ in
the denominator. For a finite system with size $L$, $t_{n}^{*}(L)$
are determined by the times at which $\lambda$ takes the local maximum.
As we shall demonstrate later, $t_{n}^{*}(L)$ does not fulfill a
simple fitting relation with $L$. Thanks to the advance of quantum
simulators, quantum simulations of DQPT were already reported in various
systems \cite{GuoXY,XueP,Monroe2017Nature,Jurcevic2017PRL,Bernien-Nature,Flaschner2018Nature,Smale,DuanLM},
such as trapped ions \cite{Monroe2017Nature,Jurcevic2017PRL}, Rydberg
atoms \cite{Bernien-Nature}, and ultracold atoms \cite{Flaschner2018Nature,Smale,DuanLM},
with finite sizes. Therefore, extracting the non-analytical signature
of DQPT in finite-size systems is important from both experimental
and theoretical aspects.

In this work, we study the non-analytical behaviors of DQPT in finite-size
systems with a twist boundary condition which can be realized by introducing
a magnetic flux $\phi_{c}$ into the periodic system. When the quench
parameter is across the equilibrium phase transition point, by tuning
the flux, we demonstrate that exact zeros of LE can be always achieved
at critical times $t_{n}^{*}$ even for a finite-size system. It is
interesting that the critical times obtained in this way are independent
of the system sizes and match exactly with the critical times obtained
in the thermodynamical limit of the corresponding periodic system.
Due to the finite size $L$, the rate function $\lambda(t)$ should
be divergent at critical times $t_{n}^{*}$, corresponding to the
exact zeros of LE. On the other hand, no exact zeros can be achieved
if there is no DQPT, when the postquench and prequench parameters
are in the same region of phases. Correspondingly, the rate function
is not sensitive to the flux and does not show any singular behavior.
Our theoretical work unveils that the emergence of the dynamical singularity
in the rate function can be viewed as a signature for detecting DQPT
in finite-size systems. Since the critical times in our theoretical
scheme are not dependent on the systems size, it provides us a convenient
way to determine the critical times in finite-size systems.

\section{Models and scheme for achieving the dynamical singularity}

To illustrate how the singularity of the rate function arises as a
result of the emergence of zero points of LE, we consider general
one-dimensional (1D) two-band systems with the Hamiltonian in momentum
space described by 
\begin{equation}
\hat{h}(\gamma,k)=\sum_{\alpha=x,y,z}d_{\alpha}(\gamma,k)\hat{\sigma}_{\alpha}+d_{0}(\gamma,k)\hat{\mathbb{I}},\label{eq:hk}
\end{equation}
where $\gamma$ denotes a phase transition driving parameter; $\hat{\sigma}_{\alpha}$
are Pauli matrices with $\alpha=x,y,z$; $d_{\alpha}(\gamma,k)$ and
$d_{0}(\gamma,k)$ are the corresponding vector components of $\hat{h}(\gamma,k)$;
and $\hat{\mathbb{I}}$ is the unit matrix. Such systems are widely
studied in the literature \cite{Dora,Budich2016PRB,YangC} and include,
e.g., the transverse-field Ising model, quantum \textit{XY} model,
the Su-Schrieffer-Heeger (SSH) model, and Creutz model, as special
cases. Consider a quench process described by a sudden change of driving
parameter $\gamma=\gamma_{i}\theta(-t)+\gamma_{f}\theta(t)$ with
the initial state prepared as the ground state of the prequench Hamiltonian
$H(\gamma_{i})$. The LE following the quench can be written as 
\begin{equation}
\mathcal{L}=\prod_{k}\mathcal{L}_{k}=\prod_{k}\left|\langle\psi_{k}^{i}|e^{-i\hat{h}(\gamma_{f},k)t}|\psi_{k}^{i}\rangle\right|^{2},
\end{equation}
where $\hat{h}(\gamma_{f},k)$ is the postquench Hamiltonian with
mode $k$. Choosing $|\psi_{k}^{i}\rangle$ as the $k$-mode of the
ground state of the prequench Hamiltonian, then we have 
\begin{equation}
\mathcal{L}_{k}=1-\Lambda_{k}\sin^{2}[\epsilon_{f}(k)t],
\end{equation}
with 
\[
\Lambda_{k}=1-\left[\frac{\sum_{\alpha=x,y,z}d_{\alpha}(\gamma_{i},k)d_{\alpha}(\gamma_{f},k)}{\epsilon_{i}(k)\epsilon_{f}(k)}\right]^{2},
\]
where $\epsilon_{i}(k)=\sqrt{\underset{\alpha}{\sum}d_{\alpha}^{2}(\gamma_{i},k)}$
and $\epsilon_{f}(k)=\sqrt{\underset{\alpha}{\sum}d_{\alpha}^{2}(\gamma_{f},k)}$.
The singularity of rate function $\lambda(t)=-\frac{1}{L}\ln\mathcal{L}(t)$
occurs when $\mathcal{L}(t)=0$, which needs at least one $k$-mode
fulfilling $\Lambda_{k}=1$ and gives rise to the following constraint
relation 
\begin{equation}
\sum_{\alpha=x,y,z}d_{\alpha}(\gamma_{i},k)d_{\alpha}(\gamma_{f},k)=0.\label{eq:dvalue}
\end{equation}
To make our discussion concrete, we consider the SSH model \cite{SSH}
and the Creutz model \cite{Creutz1999} as examples and show the details
of the calculation in this section.

\subsection{SSH model}

First, we consider the SSH model with the vector components of Hamiltonian
given by 
\begin{eqnarray}
d_{x}(k) & = & J_{1}+J_{2}\cos k,\\
d_{y}(k) & = & -J_{2}\sin k
\end{eqnarray}
and $d_{z}(k)=d_{0}(k)=0$, where $J_{1}$ and $J_{2}$ represent
the intracellular and intercellular tunneling amplitudes, respectively.
The SSH model possesses two topologically different phases for $J_{2}>J_{1}$
and $J_{2}<J_{1}$ with a phase transition occurring at the transition
point of $J_{2c}/J_{1}=1$ \cite{SSH,LiLH2014}. Then we quench parameter
$J_{2}$ from $J_{2i}$ to $J_{2f}$ at $t=0$ and get the LE of the
SSH model 
\begin{align}
\mathcal{L}(t) & =\prod_{k}\left\{ 1-\Lambda_{k}\sin^{2}[\epsilon_{f}(k)t]\right\} ,\label{Lt}
\end{align}
where $\epsilon_{f}(k)$ and $\Lambda_{k}$ are given by 
\begin{equation}
\epsilon_{f}(k)=J_{1}\sqrt{1+2\gamma_{f}\cos k+\gamma_{f}^{2}},\label{Ekf}
\end{equation}
\begin{equation}
\Lambda_{k}=1-\frac{[1+(\gamma_{i}+\gamma_{f})\cos k+\gamma_{i}\gamma_{f}]^{2}}{(1+2\gamma_{i}\cos k+\gamma_{i}^{2})(1+2\gamma_{f}\cos k+\gamma_{f}^{2})}.
\end{equation}
Here $\gamma_{i}=\frac{J_{2i}}{J_{1}}$ and$\gamma_{f}=\frac{J_{2f}}{J_{1}}$.
For convenience, we shall fix $J_{1}=1$ and take it as the energy
unit in the following calculation. For a finite size system under
the periodic boundary condition (PBC), the momentum $k$ takes discrete
values $k=2\pi m/L$ with $m=-L/2,-L/2+1,\ldots,L/2-1$ if $L$ is
even or $m=-(L-1)/2,-(L-1)/2+1,\ldots,(L-1)/2$ if $L$ is odd. 
\begin{figure}
\begin{centering}
\includegraphics[scale=0.54]{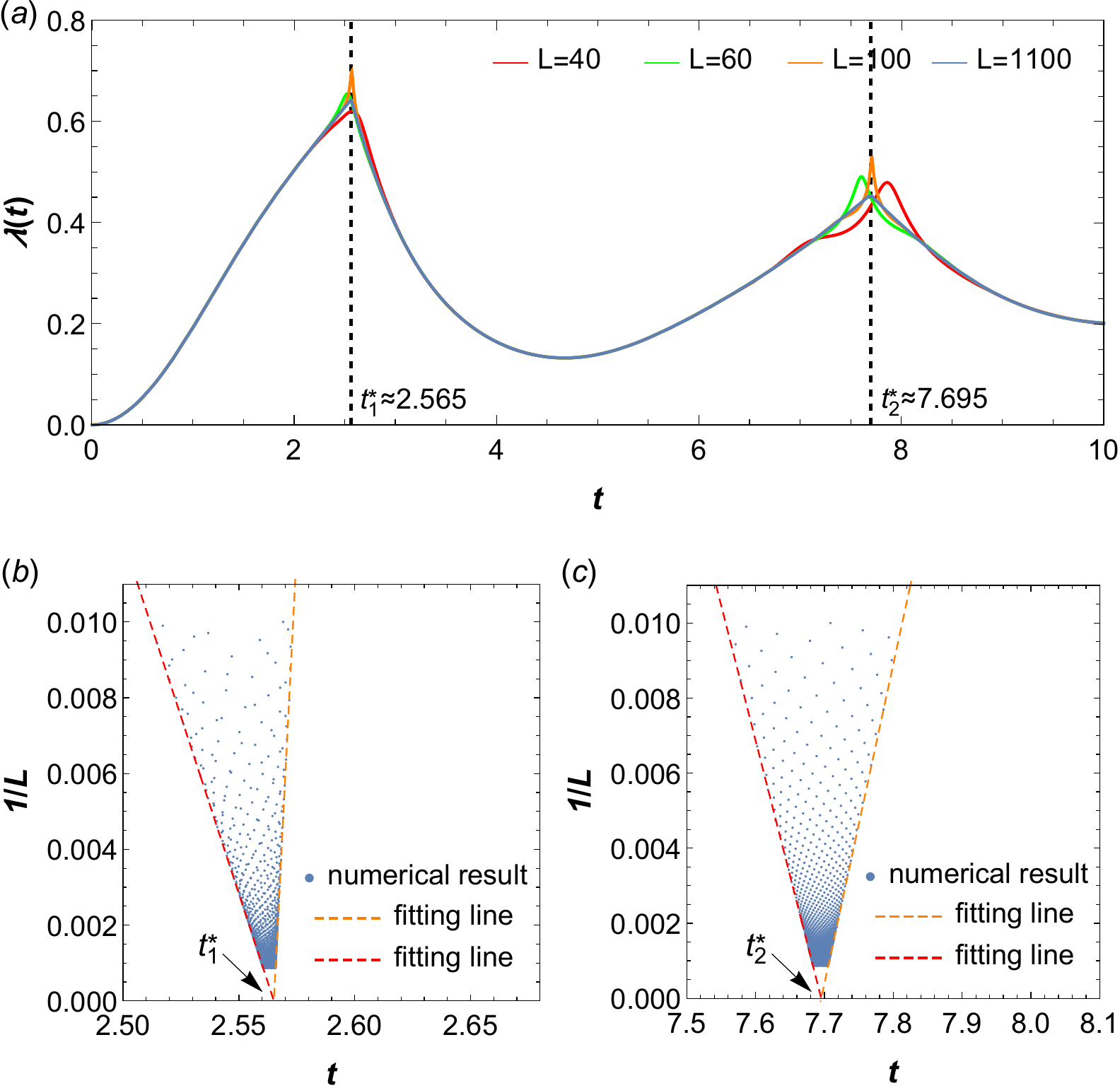} 
\par\end{centering}
\caption{(a) The rate function $\lambda(t)$ of the SSH model versus $t$ for
different system sizes $L=40$, $60$, 100 and $1100$. Vertical dashed
lines guide the values of critical times $t_{1}^{*}\approx2.565$
and $t_{2}^{*}\approx7.695$. (b), (c) are numerical results of the
time when $\lambda$ takes its local maximums for different sizes
$L$. We take $\gamma_{i}=1.5$ and $\gamma_{f}=0.5$. \label{Fig1}}
\end{figure}

For a finite-size system under the PBC, we can utilize Eqs. (\ref{rf})
and (\ref{Lt}) to calculate the rate function numerically. In Fig.
\ref{Fig1}(a), we display the rate function $\lambda(t)$ versus
time $t$ for different system sizes $L$. Around the critical times
$t_{n}^{*}$, the rate function exhibits a series of peaks and the
times $t_{n}^{*}(L)$ corresponding to these local maximums can be
used to interpolate numerically the values of critical times in the
limit of $L\rightarrow\infty$. When we increase the size, $t_{n}^{*}(L)$
does not change linearly with $L$, but approaches the critical times
$t_{n}^{*}$ in an oscillating way as shown in Figs. \ref{Fig1}(b)
and \ref{Fig1}(c). In the thermodynamical limit, the non-analytical
behaviors of $\lambda(t)$ are characterized by the emergence of a
cusp at $t_{n}^{*}$. Using $\lambda_{\mathrm{max}}$ to represent
the first local maximum of $\lambda(t)$, we find that the value of
$\lambda_{\mathrm{max}}$ does not increase linearly with the increase
of system size but approaches a finite number in an oscillating way.
Our numerical result unveils $\lambda_{\mathrm{max}}\sim0.643$ with
$L\rightarrow\infty$. In the thermodynamical limit $L\rightarrow\infty$,
the momentum $k$ distributes continuously and we have 
\[
\lambda(t)=-\frac{1}{2\pi}\int_{0}^{2\pi}\ln[1-\Lambda_{k}\sin^{2}[\epsilon_{f}(k)t]]dk,
\]
from which we numerically evaluate the value $\lambda(t_{1}^{*})\approx0.643$
at the critical time $t_{1}^{*}$. It is evident that $\lambda(t_{1}^{*})$
is equal to $\lambda_{\mathrm{max}}$ in the thermodynamical limit.
\begin{figure}
\begin{centering}
\includegraphics[scale=0.98]{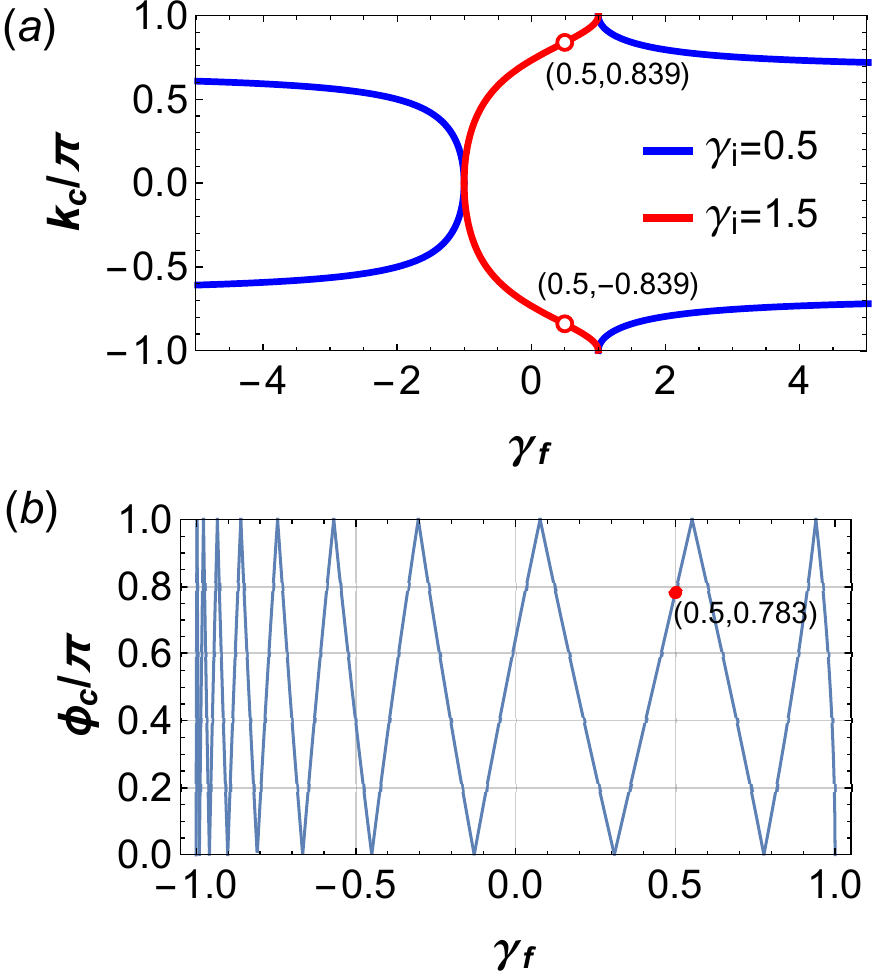} 
\par\end{centering}
\caption{(a) The images of $k_{c,+}/\pi$ and $k_{c,-}/\pi$ versus $\gamma_{f}$
for the SSH model. The blue and red lines correspond to $\gamma_{i}=0.5$
and $\gamma_{i}=1.5$, respectively. The two red circles denote $k_{c,+}/\pi\approx0.839$
and $k_{c,-}/\pi\approx-0.839$ for $\gamma_{i}=1.5$ and $\gamma_{f}=0.5$,
respectively. (b) The exact solution of $\phi_{c}/\pi$ for $\gamma_{f}\in[-1,1].$
The red point denotes $\phi_{c}/\pi\approx0.783$ for $\gamma_{f}=0.5$.
Here $\gamma_{i}=1.5$ and $L=20$. \label{Fig2}}
\end{figure}

The non-analytical behaviors of the rate function occurring at the
critical times $t_{n}^{*}$ are associated to the emergence of zeros
of LE. We notice that the constraint relation for ensuring $\mathcal{L}(t)=0$
is 
\begin{equation}
\gamma_{f}=-\frac{1+\gamma_{i}\cos k}{\gamma_{i}+\cos k}.\label{eq:gammaf}
\end{equation}
If $|\gamma_{i}|<1$, Eq. (\ref{eq:gammaf}) is fulfilled only for
$|\gamma_{f}|>1$. On the other hand, if $|\gamma_{i}|>1$, Eq. (\ref{eq:gammaf})
is fulfilled only for $|\gamma_{f}|<1$. It means that the exact zeros
of LE emerge only when the quench parameter $\gamma$ is across the
underlying phase transition point. 
When $\gamma_{i}$ and $\gamma_{f}$ are in different phase regions,
there always exists a pair of momentum modes given by 
\begin{equation}
k_{c,\pm}=\pm\arccos\left[-\frac{1+\gamma_{i}\gamma_{f}}{\gamma_{i}+\gamma_{f}}\right],\label{kc}
\end{equation}
which leads to the occurrence of a series of zero points of LE at
\begin{equation}
t_{n}^{*}=\frac{\pi}{2\epsilon_{f}(k_{c,\pm})}(2n-1),
\end{equation}
with 
\begin{equation}
\epsilon_{f}(k_{c,\pm})/J_{1}=\sqrt{\frac{(1-\gamma_{f}^{2})(\gamma_{i}-\gamma_{f})}{\gamma_{i}+\gamma_{f}}},\label{Ekc}
\end{equation}
and $n$ being a positive integer. Since $\epsilon_{f}(k_{c,+})=\epsilon_{f}(k_{c,-})$,
we omit the subscript $\pm$ in $t_{n}^{*}$ as either $\epsilon_{f}(k_{c,+})$
or $\epsilon_{f}(k_{c,-})$ gives the same contribution to critical
times. In Fig. \ref{Fig2}(a), we exhibit the images of $k_{c,+}/\pi$
and $k_{c,-}/\pi$ versus $\gamma_{f}$ for $\gamma_{i}=0.5$ and
$\gamma_{i}=1.5$ according to Eq. (\ref{kc}), and the two red circles
denote $k_{c,+}/\pi\approx0.839$ and $k_{c,-}/\pi\approx-0.839$
for $\gamma_{i}=1.5$ and $\gamma_{f}=0.5$. For finite size systems,
$k$ takes discrete values. According to Eq. (\ref{kc}), $k_{c,\pm}$
is usually not equal to the quantized momentum values $k=2\pi m/L$
enforced by the PBC except for some fine-tuned postquench parameters
\citep{ZhouBZ2021}. With the increase in the system size, $k_{c,\pm}$
can be approached in terms of $\min|k-k_{c,\pm}|\leq\pi/L$, and thus
exact zeros of LE are usually only achievable in the thermodynamical
limit of $L\rightarrow\infty$.

Although exact zeros of LE for a finite-size system generally do not
exist, next we unveil that exact zeros of LE can be achieved even
in a finite-size system if we introduce a magnetic flux $\phi$ into
the system. The effect of magnetic flux is effectively described by
the introduction of a twist boundary condition in real space $c_{L+1}^{\dagger}=c_{1}^{\dagger}e^{i\phi}(\phi\in(0,\pi])$.
Under the twist boundary condition, the quantized momentum is shifted
by a factor $\phi/L$, i.e., $k=\frac{2\pi m+\phi}{L}$ with $m=-L/2,-L/2+1,\ldots,L/2-1$
if $L$ is even or $m=-(L-1)/2,-(L-1)/2+1,\ldots,(L-1)/2$ if $L$
is odd. Therefore, for a given lattice size $L$ we can always achieve
$k_{c,+}$ or $k_{c,-}$ by tuning the flux $\phi$ to 
\begin{equation}
\phi_{c}=\min\{\!\!\negthinspace\negthinspace\!\mod[Lk_{c,+},2\pi],\negthinspace\!\!\!\!\mod[Lk_{c,-},2\pi]\}.\label{ssh-phic}
\end{equation}
In Fig. \ref{Fig2}(b), we display the image of $\phi_{c}/\pi$ versus
$\gamma_{f}$ according to Eq. (\ref{ssh-phic}) for the system with
$\gamma_{i}=1.5$ and $L=20$, and the red point in the picture denotes
$\phi_{c}/\pi\approx0.783$ for $\gamma_{i}=1.5$ and $\gamma_{f}=0.5$.

Let $\Delta=\phi-\phi_{c}$, at the time $t=t_{n}^{*}$, we can get
\begin{equation}
\lambda(t_{n}^{*})=-\frac{1}{L}[\ln\mathcal{L}_{k^{*}}(t_{n}^{*})+\sum_{k\neq k^{*}}\ln\mathcal{L}_{k}(t_{n}^{*})],
\end{equation}
where $\mathcal{L}_{k^{*}}(t_{n}^{*})$ comes from the contribution
of the $k^{*}$-mode which is closest to $k_{c}$, i.e., $k^{*}=k_{c}+\Delta/L$.
Let $\Delta\rightarrow0$, we can get 
\begin{equation}
\mathcal{L}_{k^{*}}(t_{n}^{*})\approx\frac{(\gamma_{i}+\gamma_{f})^{3}+\gamma_{f}^{2}t_{n}^{*2}(\gamma_{f}-\gamma_{i})(1-\gamma_{i}^{2})}{(\gamma_{f}-\gamma_{i})^{2}(\gamma_{f}+\gamma_{i})L^{2}}\Delta^{2}.\label{eq:eta}
\end{equation}
When $\Delta\rightarrow0$, $\mathcal{L}_{k^{*}}(t_{n}^{*})\rightarrow0$
and thus $\ln\mathcal{L}_{k^{*}}(t_{n}^{*})$ is divergent, i.e.,
when $\phi$ achieves $\phi_{c}$, the rate function becomes divergent
at the critical times. 
\begin{figure}
\begin{centering}
\includegraphics[scale=0.78]{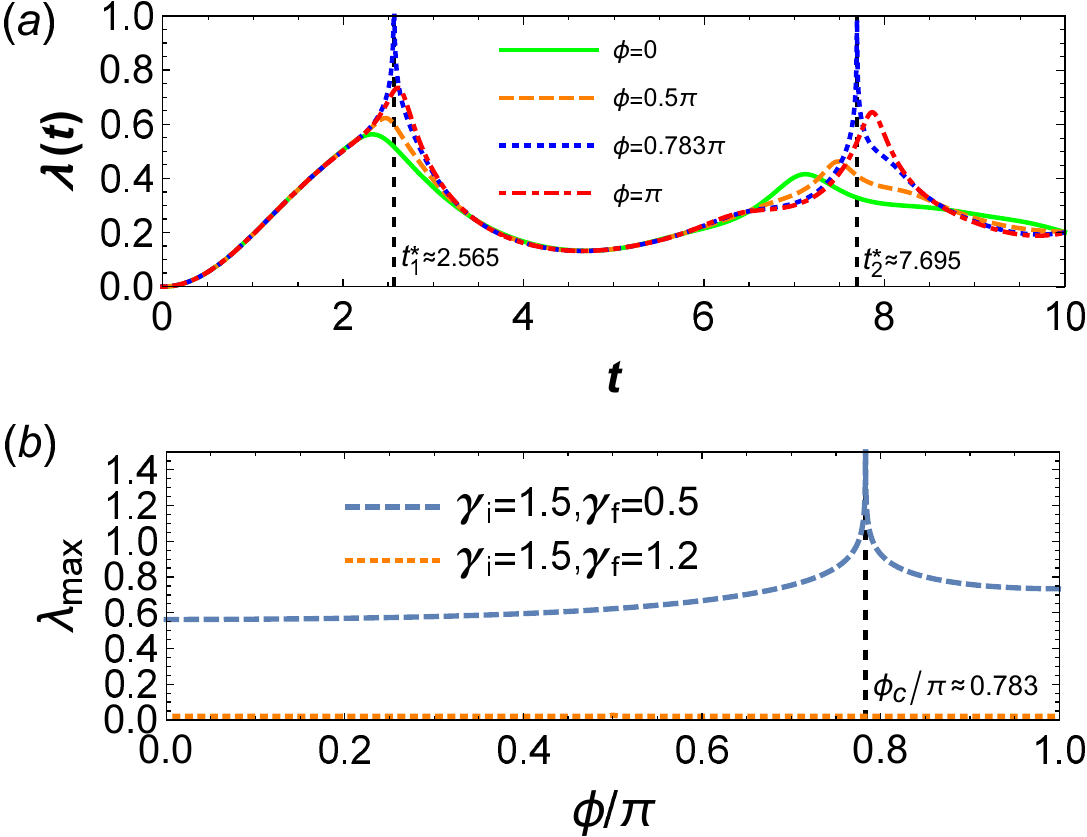} 
\par\end{centering}
\caption{(a) The rate function $\lambda(t)$ versus $t$ for the SSH model
with $\gamma_{i}=1.5,\ \gamma_{f}=0.5,\ \phi=0,0.5\pi,0.783\pi$ and
$\pi$. Vertical dashed lines guide the divergent points $t_{1}^{*}\approx2.565$
and $t_{2}^{*}\approx7.695$, respectively. (b) The images of $\lambda_{\mathrm{max}}$
versus $\phi/\pi$. The dashed blue line corresponds to $\gamma_{i}=1.5,\ \gamma_{f}=0.5$,
whereas the dotted orange line corresponds to $\gamma_{i}=1.5,\ \gamma_{f}=1.2$.
The vertical dashed line guides the divergent point $\phi_{c}/\pi\approx0.783$.
Here we take $L=20$. \label{Fig3}}
\end{figure}

In Fig. \ref{Fig3}(a), we demonstrate rate functions versus $t$
for various $\phi$ with $L=20$, $\gamma_{i}=1.5$ and $\gamma_{f}=0.5$.
It is shown that the rate function is divergent at the critical times
$t_{1}^{*}\approx2.565$ and $t_{2}^{*}\approx7.695$ when $\phi$
is tuned to the critical value $\phi_{c}$ which is shown in Fig.
\ref{Fig2}(b). In comparison with Fig. \ref{Fig1}(a), both the nonanalytical
behaviors occur at the same critical times $t_{1}^{*}$ and $t_{2}^{*}$.
While the nonanalyticity of the rate function in the thermodynamical
limit is characterized by a cusp or a kink, the nonanalyticity of
the rate function of a finite-size system induced by tuning the flux
$\phi$ is characterized by the appearance of singularity at the critical
times. Such a singularity of the rate function for the finite-size
system is a kind of dynamical singularity, which corresponds to the
occurrence of exact zeros of LE. The existence of dynamical singularity
for a finite-size system means that the initial state can evolve to
its orthogonal state at a series of time by tuning the magnetic flux.

For a given $\gamma_{i}$ and $\gamma_{f}$, tuning $\phi$ from $0$
to $\pi$, from Fig. \ref{Fig3}(b) we can see that if $\gamma_{i}$
and $\gamma_{f}$ belong to the same phase, $\lambda_{\mathrm{max}}$
barely changes with $\phi$, which means no singularity of rate function
can be observed; if $\text{\ensuremath{\gamma}}_{i}$ and $\gamma_{f}$
belong to different phases, $\lambda_{\mathrm{max}}$ will diverge
at $\phi_{c}/\pi\approx0.783$, which gives a signal of DQPT. Therefore,
we can judge whether a DQPT happens by observing the change of $\lambda_{\mathrm{max}}$
as a function of $\phi$, which continuously varies from $0$ to $\pi$.
By tuning $\phi$ in finite-size systems, we also obtain the critical
times of DQPT, which are usually defined in the thermodynamical limit
and can be extracted from the finite-size-scaling analysis in previous
studies.

\subsection{Creutz model}

Next we consider the Creutz model \cite{Creutz1999} which describes
the dynamics of a spinless electron moving in a ladder system governed
by the Hamiltonian: 
\begin{align}
H & =-\sum_{j=1}[J_{h}(e^{i\theta}c_{j+1}^{p\dagger}c_{j}^{p}+e^{-i\theta}c_{j+1}^{q\dagger}c_{j}^{q})\nonumber \\
 & \qquad\qquad+J_{d}(c_{j+1}^{p\dagger}c_{j}^{q}+c_{j+1}^{q\dagger}c_{j}^{p})+J_{v}c_{j}^{q\dagger}c_{j}^{p}+\mathrm{H.c.}],
\end{align}
where $c_{j}^{p(q)\dagger}$ and $c_{j}^{p(q)}$ are fermionic creation
and annihilation operators on the $j$th site of the lower (upper)
chain; $J_{h}$, $J_{d}$, and $J_{v}$ represent hopping amplitudes
for horizontal, diagonal, and vertical bonds, respectively; and $\theta\in[-\pi/2,\pi/2]$
represents the magnetic flux per plaquette induced by a magnetic field
piercing the ladder \citep{Jafari,Creutz1999}. Via the Fourier transformation,
the vector components of the Hamiltonian in momentum space can be
expressed as $d_{x}(k)=-2J_{d}\cos k-J_{v},\ d_{y}(k)=0,\ d_{z}(k)=-2J_{h}\sin k\sin\theta$,
and $d_{0}(k)=-2J_{h}\cos k\cos\theta$. For simplicity, in the following
we will focus on the case of $J_{h}=J_{d}=J$ and $J_{v}/2J<1$, and
take $J=1$ as the unit of energy. In this case, the Creutz model
has two distinct topologically nontrivial phases for $-\pi/2\leq\theta<0$
and $0<\theta\leq\pi/2$ with a phase transition occurring at the
transition point of $\theta=0$ \cite{LiLH}. Then we quench parameter
$\theta$ from $\theta_{i}$ to $\theta_{f}$ at $t=0$ and get the
LE of the Creutz model 
\begin{align}
\mathcal{L}(t) & =\prod_{k}\left\{ 1-\Lambda_{k}\sin^{2}[\epsilon_{f}(k)t]\right\} ,
\end{align}
where 
\begin{equation}
\Lambda_{k}=1-\frac{16J^{4}[(\cos k+\tilde{J_{v}})^{2}+\sin^{2}k\sin\theta_{i}\sin\theta_{f}]^{2}}{\epsilon_{i}^{2}(k)\epsilon_{f}^{2}(k)},
\end{equation}
\begin{equation}
\epsilon_{i}(k)=2J\sqrt{(\cos k+\tilde{J_{v}})^{2}+\sin^{2}k\sin^{2}\theta_{i}},
\end{equation}
and 
\begin{equation}
\epsilon_{f}(k)=2J\sqrt{(\cos k+\tilde{J_{v}})^{2}+\sin^{2}k\sin^{2}\theta_{f}}\label{Ekf-1}
\end{equation}
with $\tilde{J_{v}}=J_{v}/2J$. The corresponding constraint relation
of Eq. (\ref{eq:dvalue}) for the occurrence of exact zeros of LE
is 
\begin{equation}
\sin\theta_{f}=-\frac{(\cos k+\tilde{J_{v}})^{2}}{\sin^{2}k\sin\theta_{i}}.\label{eq:thetaf}
\end{equation}
If $\sin\theta_{i}<0$, Eq. (\ref{eq:thetaf}) is fulfilled only for
$\sin\theta_{f}>0$. On the other hand, if $\sin\theta_{i}>0$, Eq.
(\ref{eq:thetaf}) is fulfilled only for $\sin\theta_{f}<0$. It means
that the dynamical singularity of the rate function exists only when
the quench parameter $\theta$ is across the underlying phase transition
point.

When $\theta_{i}$ and $\theta_{f}$ are in different phase regions,
there are always two pairs of momentum modes given by 
\begin{equation}
k_{c,1\pm}=\pm\arccos\left[\frac{-\tilde{J_{v}}+\sqrt{A(\tilde{J}_{v}^{2}-1+A)}}{1-A}\right],\label{kc-1}
\end{equation}
and 
\begin{equation}
k_{c,2\pm}=\pm\arccos\left[\frac{-\tilde{J_{v}}-\sqrt{A(\tilde{J}_{v}^{2}-1+A)}}{1-A}\right],\label{kc-2}
\end{equation}
with $A=\sin\theta_{i}\sin\theta_{f}$, which lead to the occurrence
of a series of dynamical singularities of the rate function at 
\begin{equation}
t_{n,1/2}^{*}=\frac{\pi}{2\epsilon_{f}(k_{c,1\pm/2\pm})}(2n-1),
\end{equation}
with 
\begin{equation}
\epsilon_{f}(k_{c,1\pm/2\pm})/2J=\sqrt{(\sin\theta_{f}-\sin\theta_{i})\sin\theta_{f}\sin^{2}k_{c,1\pm/2\pm}}\label{Ekc-1}
\end{equation}
and $n$ being a positive integer.  Since $\epsilon_{f}(k_{c,1+/2+})=\epsilon_{f}(k_{c,1-/2-})$,
we omit the subscript $\pm$ in $t_{n,1/2}^{*}$. Similarly, $k_{c,1\pm}$
and $k_{c,2\pm}$ are usually not equal to the quantized momentum
values $k=2\pi m/L$ enforced by the PBC except for some fine-tuned
postquench parameters. It means that the exact zeros of LE of a finite-size
system generally do not exist for arbitrary $\theta_{i}$ and $\theta_{f}$.
With the increase in the system size, $k_{c,1\pm/2\pm}$ can be approached
in terms of $\min|k-k_{c,1\pm/2\pm}|\leq\pi/L$, and thus dynamical
singularities of the rate function are usually only achieved in the
limit of $L\rightarrow\infty$. 
\begin{figure}
\begin{centering}
\includegraphics[scale=0.54]{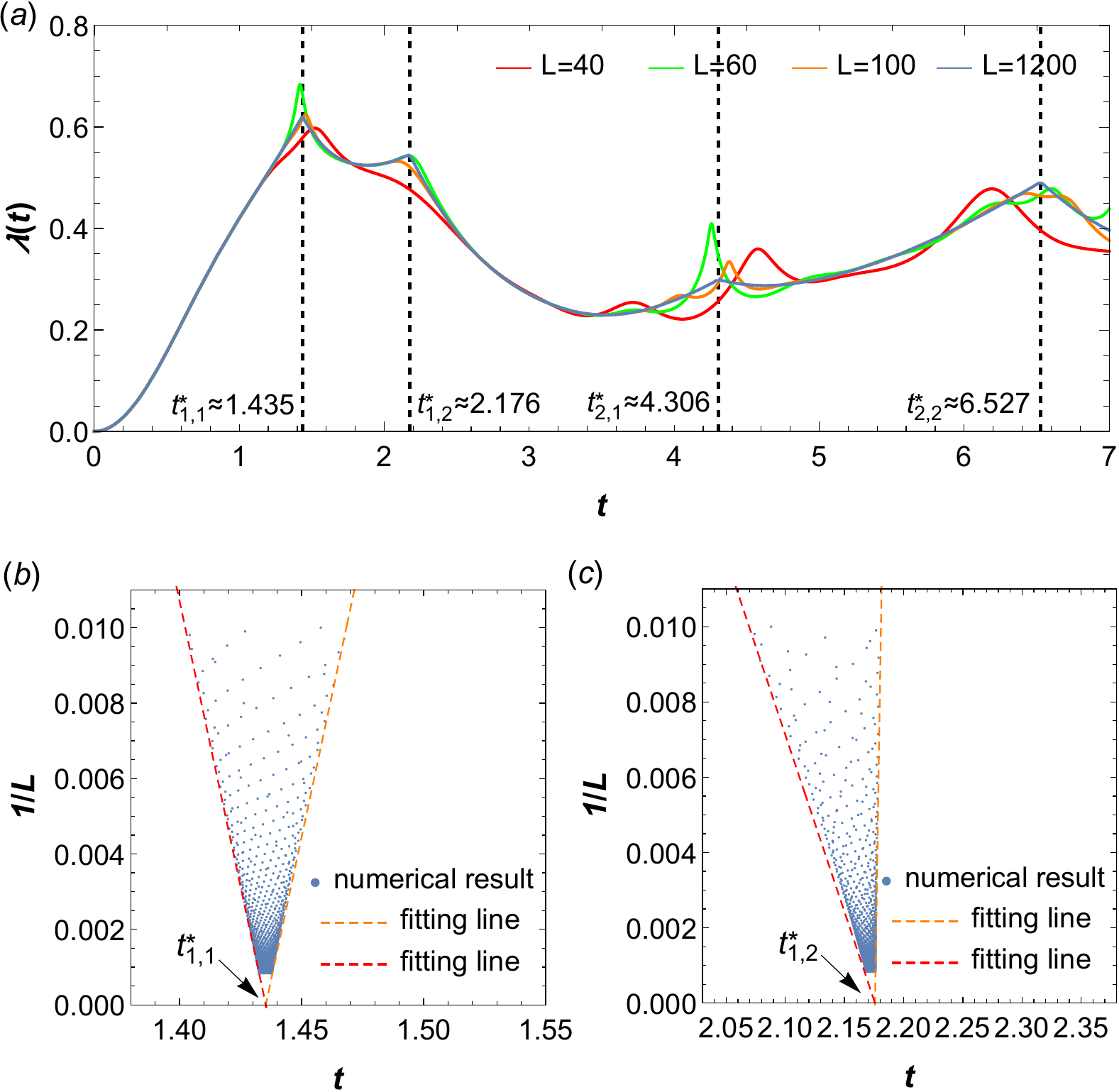} 
\par\end{centering}
\caption{(a) The rate function $\lambda(t)$ versus $t$ for the Creutz model
with different system sizes $L=40$, $60$, 100, and $1200$. Vertical
dashed lines guide the values of critical times $t_{1,1}^{*}\approx1.435$,
$t_{1,2}^{*}\approx2.176$, $t_{2,1}^{*}\approx4.306$, and $t_{2,2}^{*}\approx6.527$,
respectively. (b), (c) are numerical results of the time when $\lambda$
takes its local maximums for different sizes $L$. Here $\tilde{J}_{v}=0.5,\ \theta_{i}=0.4$,
and $\theta_{f}=-0.4$. \label{Fig4}}
\end{figure}

In Fig. \ref{Fig4}(a), we display the rate function $\lambda(t)$
versus time $t$ for different system sizes $L$. From Figs. \ref{Fig4}(b)
and \ref{Fig4}(c) we can see that $t_{1,1}^{*}(L)$ and $t_{1,2}^{*}(L)$
approach the critical times in an oscillating way as the size $L$
increases. With the increase of the size $L$, we find that the value
of $\lambda_{\mathrm{max}}$ also approaches a finite number in an
oscillating way and $\lambda_{\mathrm{max}}\sim0.621$ when $L\rightarrow\infty$.
In the thermodynamical limit, the momentum $k$ distributes continuously
and we have $\lambda(t)=-\frac{1}{2\pi}\int_{0}^{2\pi}\ln[1-\Lambda_{k}\sin^{2}[\epsilon_{f}(k)t]]dk$,
from which we numerically evaluate the value $\lambda(t_{1,1}^{*})\approx0.621$
at the critical time $t_{1,1}^{*}$, agreeing with $\lambda_{\mathrm{max}}$
in the thermodynamical limit.

\begin{figure}
\begin{centering}
\includegraphics[scale=1.01]{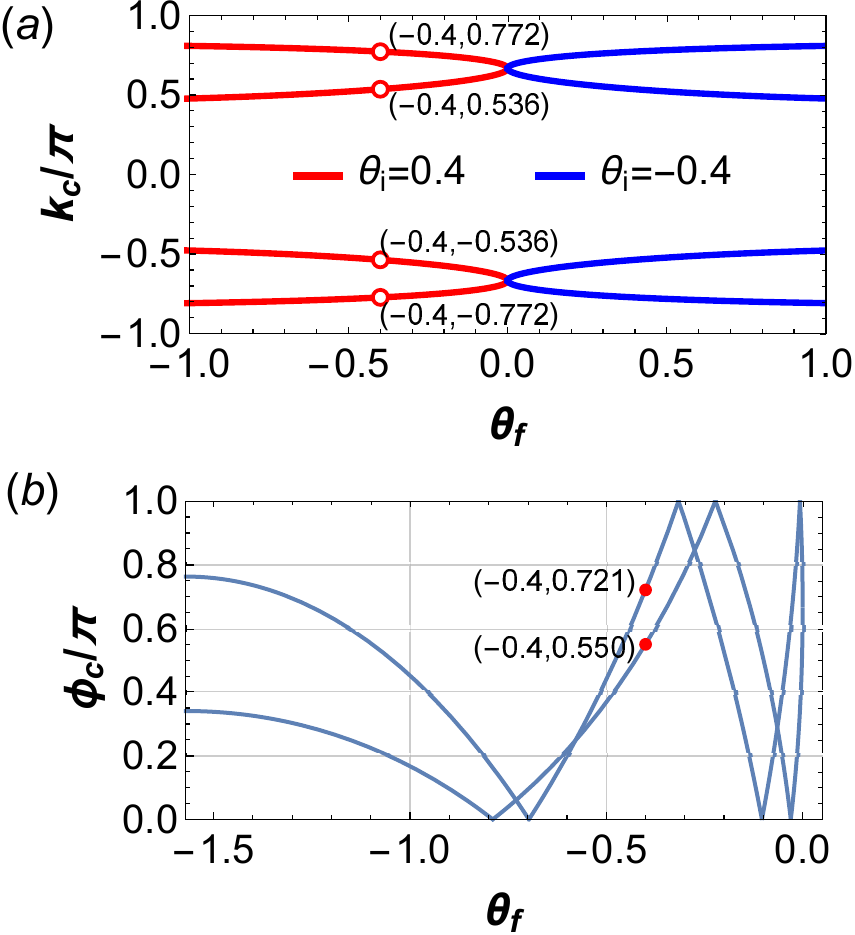} 
\par\end{centering}
\caption{(a) The images of $k_{c,1+}$, $k_{c,1-}$, $k_{c,2+}$, and $k_{c,2-}$
versus $\theta_{f}$ for the Creutz model. The blue and red lines
correspond to $\theta_{i}=-0.4$ and $\theta_{i}=0.4$, respectively.
The four red circles denote $k_{c,1\pm}/\pi\approx\pm0.536$ and $k_{c,2\pm}/\pi\approx\pm0.772$
for $\theta_{i}=0.4$ and $\theta_{f}=-0.4$. (b) The exact solutions
of $\phi_{c,1}/\pi$ and $\phi_{c,2}/\pi$ of the Creutz model for
$\theta_{f}\in[-\pi/2,0]$. The two red points denote $\phi_{c,1}/\pi\approx0.721$
and $\phi_{c,2}/\pi\approx0.550$ for $\theta_{f}=-0.4$. Here $\theta_{i}=0.4$,
$\tilde{J}_{v}=0.5$, and $L=20$. \label{Fig5}}
\end{figure}

In Fig. \ref{Fig5}(a), we exhibit the images of $k_{c,1+}$, $k_{c,1-}$,
$k_{c,2+}$, and $k_{c,2-}$ versus $\theta_{f}$ for $\theta_{i}=-0.4$
and $\theta_{i}=0.4$ according to Eqs. (\ref{kc-1}) and (\ref{kc-2}),
and the four red circles denote $k_{c,1\pm}/\pi\approx\pm0.536$ and
$k_{c,2\pm}/\pi\approx\pm0.772$ for $\theta_{i}=0.4$ and $\theta_{f}=-0.4$.
Since the quantized momenta $k$ usually do not include $k_{c,1\pm}$
and $k_{c,2\pm}$ under the PBC, we introduce the twist boundary condition
here. For a system with a given finite size $L$, we can always achieve
$k_{c,1+/2+}$ or $k_{c,1-/2-}$ by using the twist boundary condition
with 
\begin{equation}
\phi_{c,1/2}=\min\{\!\!\!\!\!\!\mod[Lk_{c,1+/2+},2\pi],\!\!\!\!\!\!\mod[Lk_{c,1-/2-},2\pi]\}.\label{Creutz-phic}
\end{equation}
Figure \ref{Fig5}(b) displays the images of $\phi_{c,1}/\pi$ and
$\phi_{c,2}/\pi$ versus $\theta_{f}$ according to Eq. (\ref{Creutz-phic})
for the system with $\theta_{i}=0.4$, $\tilde{J}_{v}=0.5$, and $L=20$,
and the two red points denote $\phi_{c,1}/\pi\approx0.721$ and $\phi_{c,2}/\pi\approx0.550$
for $\theta_{i}=0.4$ and $\theta_{f}=-0.4$.

Let $\Delta_{1/2}=\phi-\phi_{c,1/2}$, at the time $t=t_{n,1/2}^{*}$
we can get 
\begin{equation}
\lambda(t_{n,1/2}^{*})=-\frac{1}{L}\left[\ln\mathcal{L}_{k_{1/2}^{*}}(t_{n,1/2}^{*})+\sum_{k\neq k_{1/2}^{*}}\ln\mathcal{L}_{k}(t_{n,1/2}^{*})\right],
\end{equation}
where $\mathcal{L}_{k_{1/2}^{*}}(t_{n,1/2}^{*})$ comes from the contribution
of the $k_{1/2}^{*}$-mode which is closest to $k_{c,1/2}$, i.e.,
$k_{1/2}^{*}=k_{c,1/2}+\Delta_{1/2}/L$. Let $\Delta_{1/2}\rightarrow0$,
we can get 
\begin{equation}
\mathcal{L}_{k_{1/2}^{*}}(t_{n,1/2}^{*})\approx B_{1/2}\Delta_{1/2}^{2},\label{eq:eta-1}
\end{equation}
where 
\[
B_{1/2}=\frac{4[t_{n,1/2}^{*2}(\tilde{J}_{v}+\cos k_{c,1/2}\cos^{2}\theta_{f})^{2}-C]}{(\sin\theta_{f}-\sin\theta_{i})\sin\theta_{f}L^{2}},
\]
with $C=\frac{\sin\theta_{f}(\tilde{J}_{v}^{2}-1+\sin\theta_{i}\sin\theta_{f})}{\sin^{2}k_{c,1/2}(\sin\theta_{f}-\sin\theta_{i})}$.
It means when $\Delta_{1/2}\rightarrow0,$ i.e., $\phi\rightarrow\phi_{c,1/2}$,
$\mathcal{L}_{k_{1/2}^{*}}(t_{n,1/2}^{*})\propto\Delta_{1/2}^{2}$.
When $\phi$ reaches $\phi_{c,1/2}$, we can get a $k_{1/2}^{*}$-mode
which satisfies $k_{1/2}^{*}=k_{c,1/2}$ and $\mathcal{L}_{k_{1/2}^{*}}(t_{n,1/2}^{*})=0$,
thus the rate function is divergent at $t_{n,1/2}^{*}$. 
\begin{figure}
\begin{centering}
\includegraphics[scale=0.43]{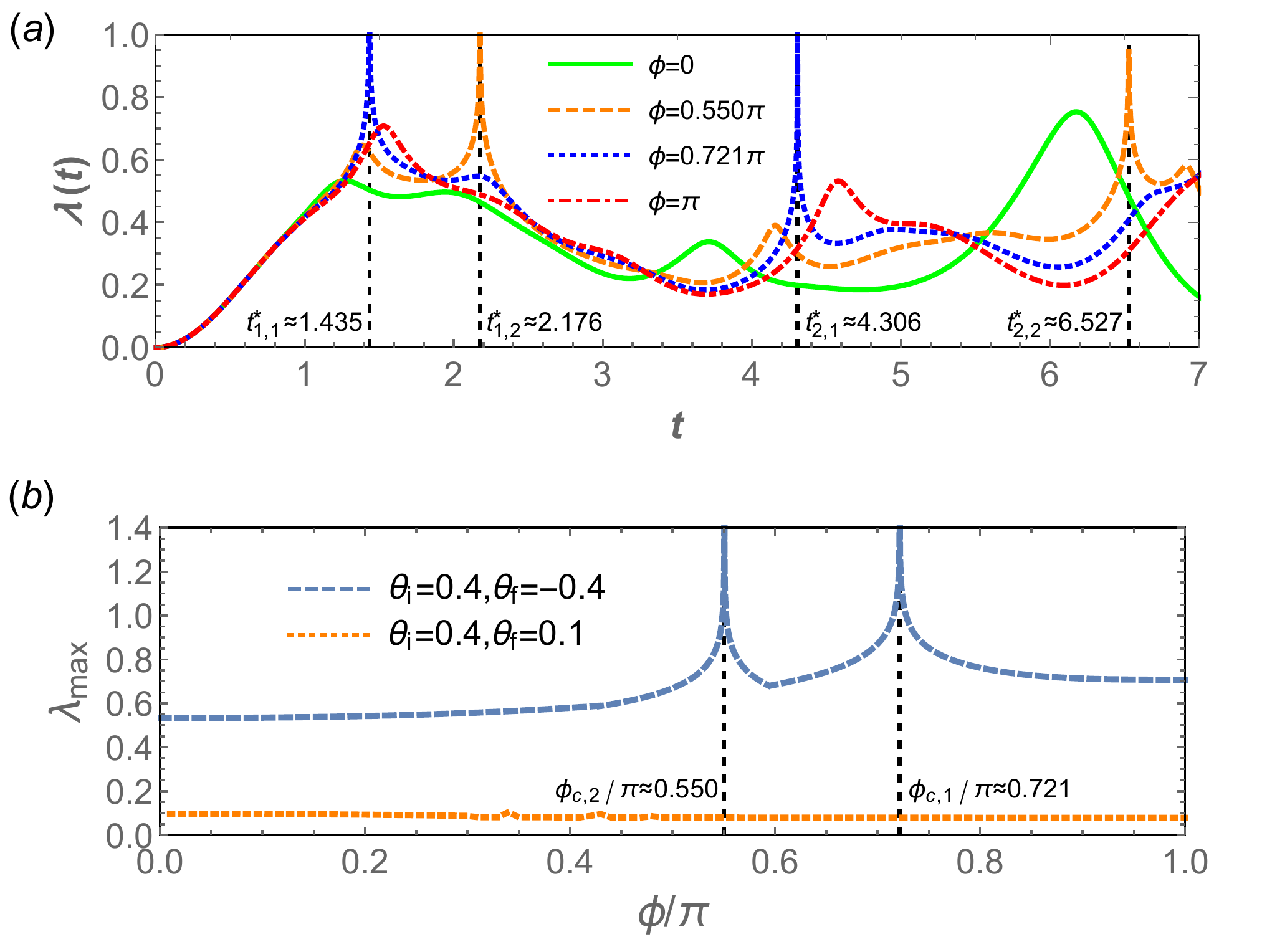} 
\par\end{centering}
\caption{(a) The rate function $\lambda(t)$ versus $t$ for the Creutz model
with $\theta_{i}=0.4,\ \theta_{f}=-0.4,\ \phi=0,0.550\pi,0.721\pi$,
and $\pi$, respectively. Vertical dashed lines guide the divergent
points $t_{1,1}^{*}\approx1.435$, $t_{1,2}^{*}\approx2.176$, $t_{2,1}^{*}\approx4.306$,
and $t_{2,2}^{*}\approx6.527$. (b) The images of $\lambda_{\mathrm{max}}$
versus $\phi/\pi$. The dashed blue line corresponds to $\theta_{i}=0.4,\ \theta_{f}=-0.4$,
and the dotted orange line corresponds to $\theta_{i}=0.4,\ \theta_{f}=0.1$.
Vertical dashed lines guide the divergent points $\phi_{c,1}/\pi\approx0.721$
and $\phi_{c,2}/\pi\approx0.550$. Here we take $\tilde{J}_{v}=0.5$
and $L=20$. \label{Fig6}}
\end{figure}

In Fig. \ref{Fig6}(a), we demonstrate rate functions versus $t$
for various $\phi$ with $L=20$, $\tilde{J}_{v}=0.5$, $\theta_{i}=0.4$,
and $\theta_{f}=-0.4$. It is shown that the rate functions are divergent
at the critical times $t_{1,1}^{*}\approx1.435$ and $t_{2,1}^{*}\approx4.306$,
when $\phi$ is tuned to the critical value $\phi_{c,1}\approx0.721\pi$,
and divergent at the critical times $t_{1,2}^{*}\approx2.176$ and
$t_{2,2}^{*}\approx6.527$, when $\phi$ is tuned to the critical
value $\phi_{c,2}\approx0.550\pi$. In comparison with Fig. \ref{Fig4}(a),
all the nonanalytical behaviors occur at the same critical times obtained
by finite-size-scaling analysis. For a pair of given $\theta_{i}$
and $\theta_{f}$, Fig. \ref{Fig6}(b) shows that if $\theta_{i}$
and $\theta_{f}$ belong to the same phase, $\lambda_{\mathrm{max}}$
only changes slightly with $\phi$, which means the absence of DQPT;
if $\theta_{i}$ and $\theta_{f}$ belong to different phases, $\lambda_{\mathrm{max}}$
will diverge at $\phi_{c,1}/\pi$ and $\phi_{c,2}/\pi$, indicating
the occurrence of DQPT. It is a remarkable fact that there are two
critical magnetic fluxes $\phi_{c,1}$ and $\phi_{c,2}$, which are
generated by two pairs of momentum modes $k_{c,1\pm}$ and $k_{c,2\pm}$,
respectively. While $\phi_{c,1}$ only produces singularities at $t_{n,1}^{*}$,
$\phi_{c,2}$ produces singularities at $t_{n,2}^{*}$.

\section{Application to other model systems}

To exhibit the applicability of our theoretical scheme to more general
cases, here we study more examples by taking account into the effect
long-range hopping, dimensionality, and interaction. We shall explore
the dynamical singularity of rate function in the SSH model with long-range
hopping, the two-dimensional Qi-Wu-Zhang model \citep{QiWuZhang2006PRB},
and the interacting SSH model, respectively.

\subsection{SSH model with long-range hopping}

\begin{figure}
\begin{centering}
\includegraphics[scale=0.52]{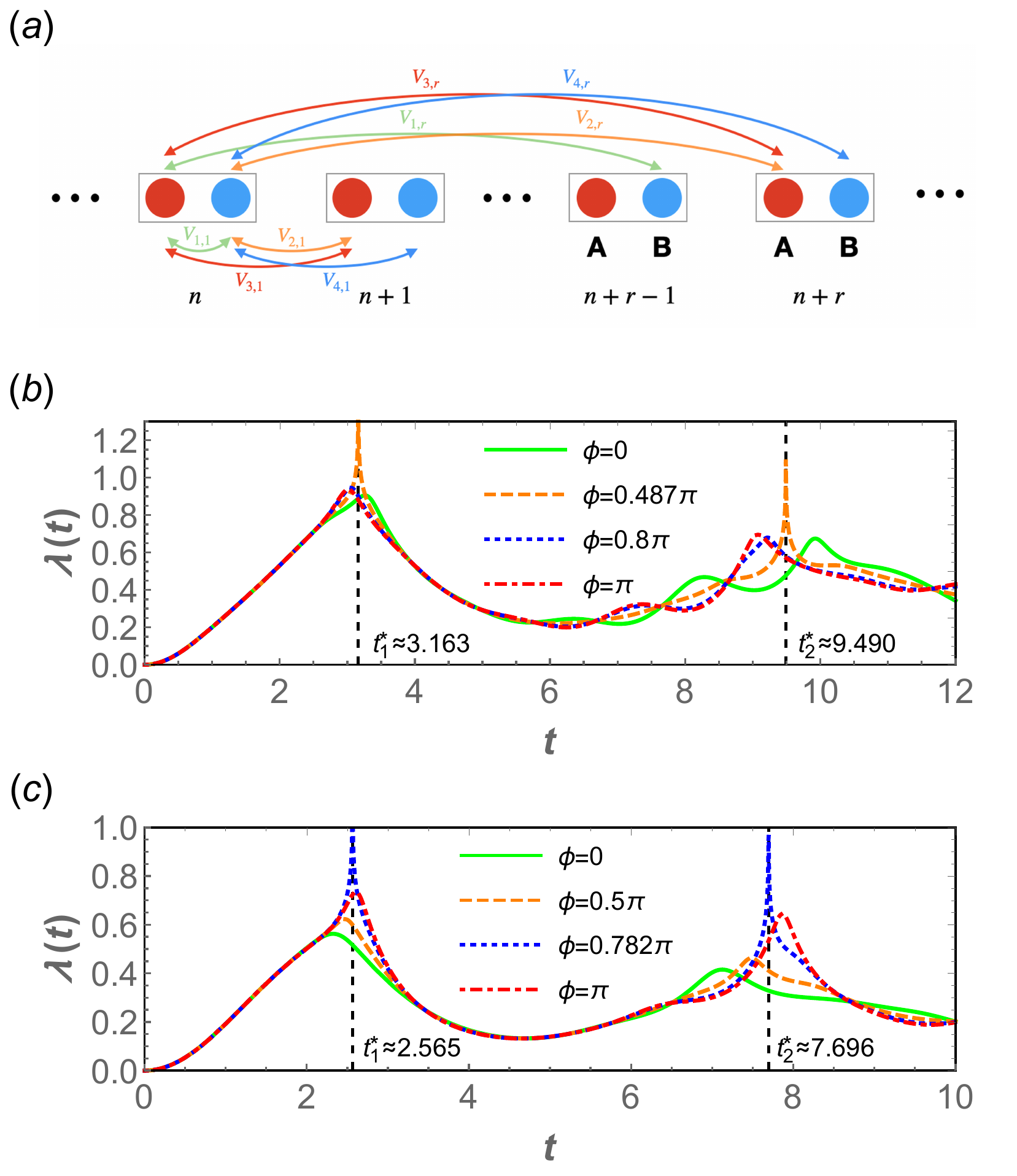} 
\par\end{centering}
\caption{(a) A scheme for the SSH model with long-range hopping. (b) The rate
function $\lambda(t)$ versus $t$ with $\phi=0,0.487\pi,0.8\pi$,
and $\pi$ for the system with $\alpha=1$, respectively. Vertical
dashed lines guide the divergent points $t_{1}^{*}\approx3.163$ and
$t_{2}^{*}\approx9.490$. (c) The rate function $\lambda(t)$ versus
$t$ with $\phi=0,0.5\pi,0.782\pi$, and $\pi$ for the system with
$\alpha=10$, respectively. Vertical dashed lines guide the divergent
points $t_{1}^{*}\approx{2.565}$ and $t_{2}^{*}\approx{7.696}$.
Here we take $J_{1}=1,\ J_{2i}=1.5,\ J_{2f}=0.5$, and $L=20$. }
\label{fig:long-range} 
\end{figure}

Consider the SSH model with long-range hopping described by 
\begin{align}
H & =\sum_{n=1}^{L}\sum_{r=1}^{L/2}(V_{1,r}c_{A,n}^{\dagger}c_{B,n+r-1}+V_{2,r}c_{A,n+r}^{\dagger}c_{B,n}\nonumber \\
 & +V_{3,r}c_{A,n}^{\dagger}c_{A,n+r}+V_{4,r}c_{B,n}^{\dagger}c_{B,n+r}+\mathrm{H.c.}),
\end{align}
where $V_{1,r}=J_{1}e^{-\alpha(r-1)},\ V_{2,r}=J_{2}e^{-\alpha(r-1)},\ V_{3,r}=J_{3}e^{-\alpha(r-1)},\ V_{4,r}=J_{4}e^{-\alpha(r-1)}$,
and $\alpha$ is a tunable positive parameter. The model is schematically
depicted in Fig. \ref{fig:long-range}(a). By using the Fourier transformation
$c_{A/B,n}^{\dagger}=\frac{1}{\sqrt{L}}\sum_{k}e^{ikn}c_{A/B,k}^{\dagger}$,
we get 
\begin{align}
H & =\sum_{k}\sum_{r=1}^{L/2}[(V_{1,r}e^{-ik(r-1)}+V_{2,r}e^{ikr})c_{A,k}^{\dagger}c_{B,k}\nonumber \\
 & +\cos[kr](V_{3,r}c_{A,k}^{\dagger}c_{A,k}+V_{4,r}c_{B,k}^{\dagger}c_{B,k})+\mathrm{H.c.}].
\end{align}

The vector components of the Hamiltonian in momentum space are 
\begin{align}
d_{x}(k) & =\sum_{r=1}^{L/2}(V_{1,r}\cos[k(r-1)]+V_{2,r}\cos[kr]),\\
d_{y}(k) & =\sum_{r=1}^{L/2}(V_{1,r}\sin[k(r-1)]-V_{2,r}\sin[kr]),\\
d_{z}(k) & =\sum_{r=1}^{L/2}(V_{3,r}-V_{4,r})\cos[kr],\\
d_{0}(k) & =\sum_{r=1}^{L/2}(V_{3,r}+V_{4,r})\cos[kr].
\end{align}
For simplicity, we set $V_{3,r}=V_{4,r}$ and choose $V_{2,r}$ as
the quench parameter. The phase transition point is $V_{2c,r}/V_{1,r}=1$.
According to Eq. (\ref{eq:dvalue}), the corresponding constraint
relation for the occurrence of divergence of the rate function is
\begin{align}
\sum_{r=1}^{L/2}(V_{1,r}\cos[k(r-1)]+V_{2i,r}\cos[kr])\nonumber \\
\times\sum_{r=1}^{L/2}(V_{1,r}\cos[k(r-1)]+V_{2f,r}\cos[kr])\nonumber \\
+\sum_{r=1}^{L/2}(V_{1,r}\sin[k(r-1)]-V_{2i,r}\sin[kr])\nonumber \\
\times\sum_{r=1}^{L/2}(V_{1,r}\sin[k(r-1)]-V_{2f,r}\sin[kr]) & =0.\label{eq:kc}
\end{align}

If we choose $J_{1}=1,\ J_{2i}=1.5,\ J_{2f}=0.5,\ \alpha=1$, and
$L=20$, we can get $k_{c}\approx0.676\pi$ from Eq. (\ref{eq:kc}).
Then it follows that $\phi_{c}\approx0.487\pi$, $t_{1}^{*}\approx3.163$,
and $t_{2}^{*}\approx9.490$. In Fig. \ref{fig:long-range}(b), we
show the image of rate function for the case of $\alpha=1$ with various
$\phi$. It is obvious that the rate function diverges at $t_{1}^{*}$
and $t_{2}^{*}$ for $\phi_{c}\approx0.487\pi$, while the rate function
is analytic for other values of $\phi$. As a comparison, we choose
$\alpha=10$ and keep the other parameters the same as the case of
$\alpha=1$, and the rate function is displayed in Fig. \ref{fig:long-range}(c).
In this case, the amplitude of hopping decays rapidly so that the
dynamical behavior of the model resembles the SSH model without long-range
hopping. Similar to Fig. \ref{Fig3}(a), the result for the case of
$\alpha=10$ in Fig. \ref{fig:long-range}(c) shows that the rate
function diverges at $t_{1}^{*}\approx{2.565}$ and $t_{2}^{*}\approx{7.696}$
for $\phi_{c}\approx0.782\pi$.

\subsection{Qi-Wu-Zhang model}

Next we consider the Qi-Wu-Zhang model, which is a two-dimensional
two-band model described by 
\begin{align}
H & =-\frac{1}{2}\sum_{n_{x},n_{y}}[(c_{n_{x},n_{y}}^{\dagger}c_{n_{x}+1,n{}_{y}}+c_{n_{x},n_{y}}^{\dagger}c_{n_{x},n{}_{y}+1})\nonumber \\
 & +ic_{n_{x},n_{y}}^{\dagger}c_{n_{x},n{}_{y}+1}^{\dagger}-c_{n_{x},n_{y}}^{\dagger}c_{n_{x}+1,n{}_{y}}^{\dagger}\nonumber \\
 & +\mu c_{n_{x},n_{y}}^{\dagger}c_{n_{x},n{}_{y}}+\mathrm{H.c.}],
\end{align}
where $\mu$ is the chemical potential. After the Fourier transformation,
we get the vector components of the Hamiltonian in momentum space
as 
\begin{align}
d_{x} & =\sin k_{y},\\
d_{y} & =-\sin k_{x},\\
d_{z} & =-\cos k_{x}-\cos k_{y}-\mu,\\
d_{0} & =-2\mu.
\end{align}
Depending on the value of $\mu$, the Qi-Wu-Zhang model is known to
have three different topological phases characterized by different
band Chern numbers with transition points at $\mu_{c}=0$ and $\pm2$
\citep{QiWuZhang2006PRB}.

According to Eq. (\ref{eq:dvalue}), the corresponding constraint
relation for the occurrence of divergence of the rate function is
\begin{equation}
\mu_{i}\mu_{f}+(\cos k_{x}+\cos k_{y})(\mu_{i}+\mu_{f})+2\cos k_{x}\cos k_{y}+2=0.\label{eq:2D}
\end{equation}
For the Qi-Wu-Zhang model, we find plenty of pairs of $(k_{xc},k_{yc})$
satisfy Eq. (\ref{eq:2D}), and the value of $k_{yc}$ is determined
by 
\begin{equation}
k_{yc}=\pm\arccos\left[\frac{-\mu_{i}\mu_{f}-(\mu_{i}+\mu_{f})\cos k_{xc}-2}{\mu_{i}+\mu_{f}+2\cos k_{xc}}\right].
\end{equation}
Taking $\mu$ as the quench parameter, we choose $\mu_{i}\in(0,2)$
and $\mu_{f}\in(-2,0)$. To make sure that $k_{yc}$ is real, the
value of $k_{xc}$ is bounded as follows: 
\begin{align}
\cos k_{xc}\in & \left[-1,-\frac{\mu_{i}\mu_{f}}{\mu_{i}+\mu_{f}+2}-1\right]\nonumber \\
 & \cup\left[-\frac{\mu_{i}\mu_{f}}{\mu_{i}+\mu_{f}-2}+1,1\right].
\end{align}
It should be noted that the critical time $t_{n}^{*}$ is no longer
a value, but in a region: 
\begin{equation}
t_{n}^{*}\in\left[\frac{(2n-1)\pi}{2\sqrt{\frac{\mu_{f}(\mu_{f}-2)(\mu_{f}-\mu_{i})}{\mu_{i}+\mu_{f}-2}}},\frac{(2n-1)\pi}{2\sqrt{\frac{\mu_{f}(\mu_{f}+2)(\mu_{f}-\mu_{i})}{\mu_{i}+\mu_{f}+2}}}\right],
\end{equation}
where $n$ is a positive integer. 
\begin{figure}
\begin{centering}
\includegraphics[scale=0.46]{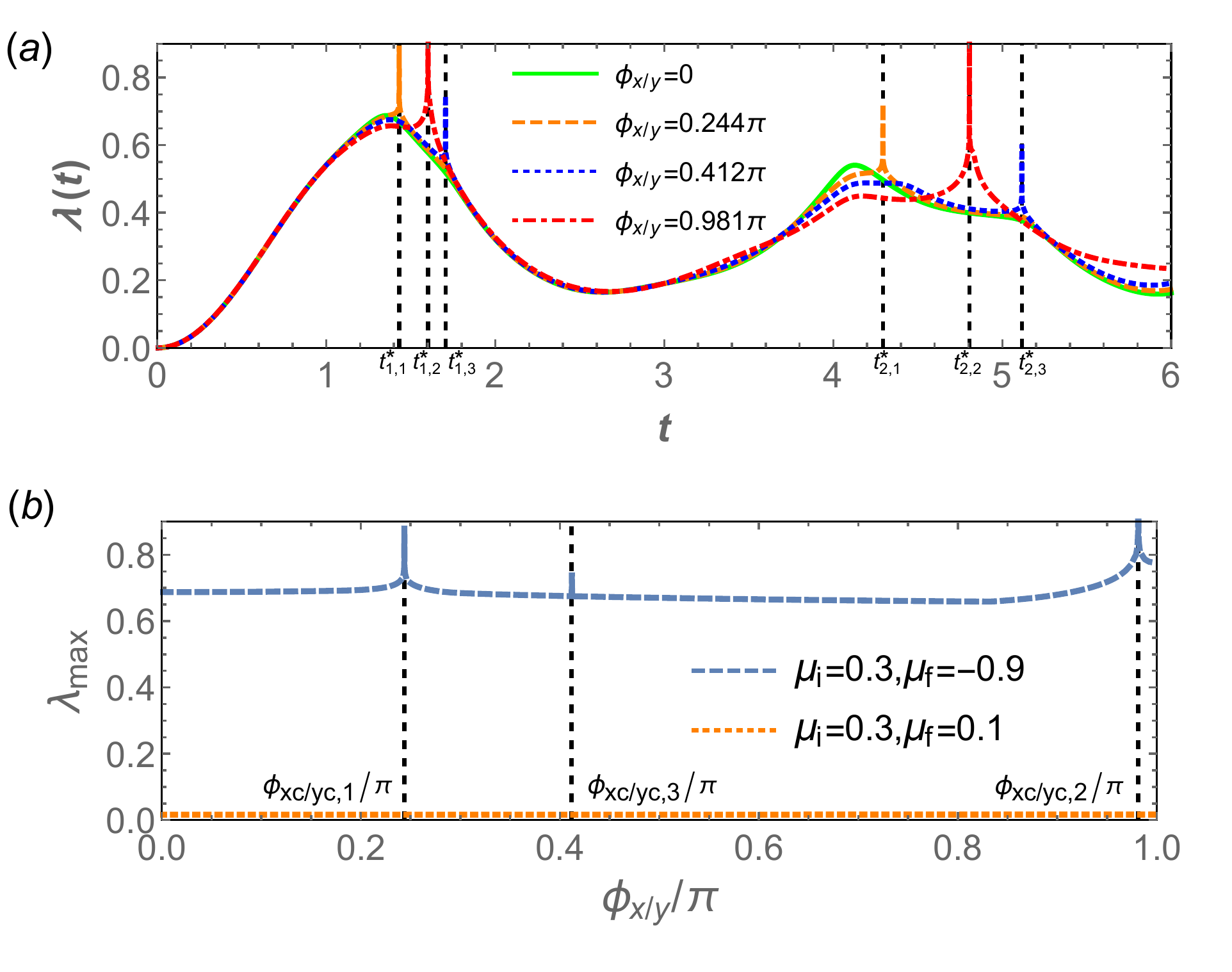} 
\par\end{centering}
\caption{(a) The rate function $\lambda(t)$ versus $t$ for the Qi-Wu-Zhang
model with $\mu_{i}=0.3,\ \mu_{f}=-0.9,$ $\phi_{x/y}=0,\ 0.244\pi,\ 0.412\pi$,
and $0.981\pi$, respectively. Vertical dashed lines guide the divergent
points $t_{1,1}^{*}\approx1.431,\ t_{1,2}^{*}\approx1.602,\ t_{1,3}^{*}\approx1.705,\ t_{2,1}^{*}\approx4.294,\ t_{2,2}^{*}\approx4.805$,
and $t_{2,3}^{*}\approx5.116$. (b) The images of $\lambda_{\mathrm{max}}$
versus $\phi_{x/y}/\pi$. The dashed blue line corresponds to $\mu_{i}=0.3,\ \mu_{f}=-0.9$,
and the dotted orange line corresponds to $\mu_{i}=0.3,\ \mu_{f}=0.1$.
Vertical dashed lines guide the divergent points $\phi_{xc/yc,1}/\pi\approx0.244,\ \phi_{xc/yc,3}/\pi\approx0.412$,
and $\phi_{xc/yc,2}/\pi\approx0.981$. Here we take $L_{x}=L_{y}=12$.
\label{fig:Qi-Wu-Zhang}}
\end{figure}

It is worth noting that, for a finite-size system, both $k_{x}$ and
$k_{y}$ take discrete values. Thus the number of pairs of $(k_{xc},k_{yc})$
satisfing Eq. (\ref{eq:2D}) is finite. If we choose $\mu_{i}=0.3,\ \mu_{f}=-0.9$,
and $L_{x}=L_{y}=12$, we can get 16 pairs of $(k_{xc},k_{yc})$ from
Eq. (\ref{eq:2D}), given by $(k_{xc,1},k_{yc,1})\approx(\pm0.146\pi,\pi)$
or $(\pi,\pm0.146\pi),$ $(k_{xc,2},k_{yc,2})\approx(\pm0.0849\pi,\pm5\pi/6)$
or $(\pm5\pi/6,\pm0.0849\pi)$, and $(k_{xc,3},k_{yc,3})\approx(\pm0.799\pi,0)$
or $(0,\pm0.799\pi)$. By applying the twist boundary conditions $(k_{x}=\frac{2\pi m_{x}+\phi_{xc}}{L_{x}},k_{y}=\frac{2\pi m_{y}}{L_{y}})$
or $(k_{x}=\frac{2\pi m_{x}}{L_{x}},k_{y}=\frac{2\pi m_{y}+\phi_{yc}}{L_{y}})$
with $\phi_{xc/yc,1}\approx0.244\pi,$ $\phi_{xc/yc,2}\approx0.981\pi$,
and $\phi_{xc/yc,3}\approx0.412\pi$, we get the corresponding critical
times given by $t_{n,1}^{*}\approx1.431(2n-1),\ t_{n,2}^{*}\approx1.602(2n-1)$,
and $t_{n,3}^{*}\approx1.705(2n-1)$, respectively. In Fig. \ref{fig:Qi-Wu-Zhang}(a),
we show the image of the rate function for the Qi-Wu-Zhang model for
various $\phi_{x/y}$, where $\mu_{i}=0.3,\ \mu_{f}=-0.9$ and $L_{x}=L_{y}=12$.
The rate functions distinctly diverge at the corresponding critical
times for $\phi_{x/y}=\phi_{xc/yc,1},\ \phi_{x/y}=\phi_{xc/yc,2}$,
and $\phi_{x/y}=\phi_{xc/yc,3}$, while the rate function under the
period boundary condition is smooth for all time. Figure \ref{fig:Qi-Wu-Zhang}(b)
exhibits that $\lambda_{\mathrm{max}}$ diverges at $\phi_{xc/yc,1}\approx0.244\pi,$
$\phi_{xc/yc,2}\approx0.981\pi$, and $\phi_{xc/yc,3}\approx0.412\pi$
when $\mu_{i}$ and $\mu_{f}$ are in different phase regions, while
there is no divergence when $\mu_{i}$ and $\mu_{f}$ belong to the
same phase. It is important to note that each $\phi_{xc/yc}$ just
produces a part of all the singularities. To obtain the whole critical
times $t_{1}^{*}\in[1.431,1.705]$, $t_{2}^{*}\in[4.294,5.116]$,
and so on, one needs to add proper twist boundary conditions on both
the $x$ and $y$ directions simultaneously.

\subsection{Interacting SSH model}

To check whether our theoretical scheme works for the interacting
system, now we consider the interacting SSH model with the twist boundary
condition, 
\begin{align}
H= & \sum_{j=1}^{L-1}\left(J_{1}c_{j,A}^{\dagger}c_{j,B}+J_{2}c_{j,B}^{\dagger}c_{j+1,A}+\mathrm{H.c.}\right)\nonumber \\
 & +(J_{1}c_{L,A}^{\dagger}c_{L,B}+J_{2}e^{-i\phi}c_{L,B}^{\dagger}c_{1,A}+\mathrm{H.c.})\nonumber \\
 & +U\sum_{j=1}^{L}\left(n_{j,A}n_{j,B}+n_{j,B}n_{j+1,A}\right),
\end{align}
where $U>0$ characterizes the strength of the nearest-neighbor repulsive
interaction, $n_{j,A/B}$ denotes the fermion occupation number operator
of sublattice $A/B$ on the unit cell $j$, and $n_{L+1,A}=n_{1,A}$.
Here we shall consider the half-filling case. The topological phase
transition of the interacting SSH model under the periodic boundary
condition was discussed in Ref. \cite{Tang}. When $U$ is much larger
than $J_{1}$ and $J_{2}$, the system is in a density-wave phase
with the ground state approximately described by $|1010\cdots\rangle$
or $|0101\cdots\rangle$. Here we shall consider the case with $U$
much smaller than $J_{1}$ and $J_{2}$, for which there is still
a phase transition when we change the parameter $J_{2}/J_{1}$ with
the transition point close to $J_{2}/J_{1}=1$ when $U$ is small.

\begin{figure}
\begin{centering}
\includegraphics[scale=0.22]{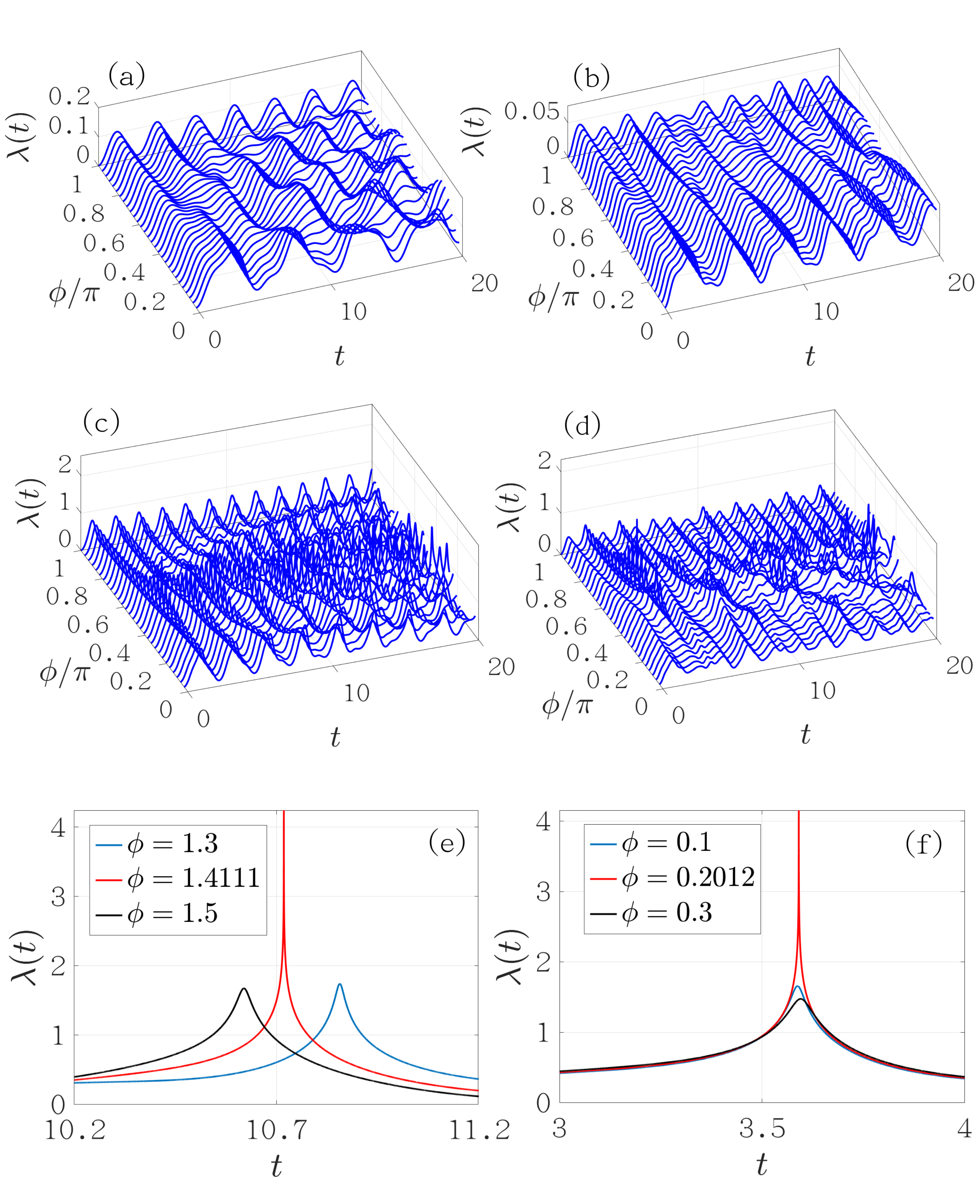} 
\par\end{centering}
\caption{The rate function $\lambda(t)$ versus $t$ for the interacting SSH
model with $J_{1}=1,J_{2i}=0.2$, and $L=5$. The other parameters
are (a) $U=0.1$, $J_{2f}=0.7$; (b) $U=0.6$, $J_{2f}=0.7$; (c),(e)
$U=0.1$, $J_{2f}=2$; (d),(f) $U=0.6$, $J_{2f}=2$. \label{fig:issh}}
\end{figure}

Since our motivation is to observe the signature of dynamical singularity
in a small size system, we shall not pursue determining the phase
boundary of the system precisely. By applying the finite-size scaling
of fidelity \cite{Tang}, we obtain the approximate value of the phase
transition point $J_{2c}/J_{1}\approx1.038$ and $J_{2c}/J_{1}\approx1.103$
for $U=0.1$ and $U=0.6$, respectively. We numerically calculate
the rate function by exact diagonalization of a system with $L=5$
by fixing $U$ and $J_{1}=1$ and quenching the parameter $J_{2}$.
The numerical results are shown in Fig. \ref{fig:issh}. For quench
from $J_{2}/J_{1}<1$ to $J_{2}/J_{1}>1$, we can observe that the
rate functions present some peaks at some typical magnetic fluxes
$\phi$, as shown in Figs. \ref{fig:issh}(c) and \ref{fig:issh}(d).
However, these peaks are absent for the quench process from $J_{2}/J_{1}<1$
to $J_{2}/J_{1}<1$, as shown in Figs. \ref{fig:issh}(a) and \ref{fig:issh}(b).
By scrutinizing these peaks in Figs. \ref{fig:issh}(c) and \ref{fig:issh}(d),
we choose three rate functions from Figs. \ref{fig:issh}(c) and \ref{fig:issh}(d)
and show them in Figs. \ref{fig:issh}(e) and \ref{fig:issh}(f),
respectively. We can see that the rate function with $\phi=1.4111$
in Fig. \ref{fig:issh}(e) and the rate function with $\phi=0.2012$
in Fig. \ref{fig:issh}(f) exhibit obvious peaks, while the other
four rate functions display no divergence. Our numerical results indicate
a clear signature of dynamical singularity even in a small size interacting
system by tuning the magnetic flux.

\section{Conclusion and discussion}

In summary, we proposed a theoretical scheme for studying the dynamical
singularity of the rate function in finite-size quantum systems which
exhibit DQPT in the thermodynamic limit. The dynamical singularity
of the rate function occurs whenever the corresponding LE has exact
zero points, which is, however, not accessible in a finite-size quantum
system with the PBC because the momentum takes quantized values $k=2\pi m/L$.
To realize the exact zeros of LE, we consider the twist boundary condition
by applying a magnetic flux into the system, which enables us to shift
the quantized momentum continuously to achieve the exact zeros of
LE. Taking the SSH model and Creutz model as concrete examples, we
demonstrate that tuning the magnetic flux can lead to the occurrence
of divergency in the rate function of a finite-size system at the
same critical times as in the case of the thermodynamical limit, when
the quench parameter is across the underlying equilibrium phase transition
point. We also exhibit the applicability of our theoretical scheme
to more general cases, including the SSH model with long-range hopping,
the Qi-Wu-Zhang model, and the interacting SSH model.

Our work unveils that the singularity of the rate function is accessible
in finite-size quantum systems by introducing an additional magnetic
flux, which provides a possible way for experimentally detecting DQPT
and the critical times in finite-size quantum systems. For the experimental
setup in a trapped-ion quantum simulator \cite{Monroe2017Nature,Jurcevic2017PRL},
it remains a challenging task to create tunable magnetic flux in the
setup. However, for the cold atomic system, one can implement discrete
momentum states by using multifrequency Bragg lasers to realize the
SSH model on a momentum lattice \cite{YanB}, where the synthetic
magnetic flux through the ring are tunable \cite{YanB2020}. We expect
that the momentum lattice of cold atomic systems might be a promising
platform to observe the dynamical singularity of the rate function
related to the DQPT. 
\begin{acknowledgments}
The work is supported by the National Key Research and Development
Program of China (Grant No. 2021YFA1402104), the NSFC under Grants
No.12174436 
and No.T2121001, and the Strategic Priority Research Program of Chinese
Academy of Sciences under Grant No. XDB33000000. 
\end{acknowledgments}

\end{document}